\documentclass[a4paper]{article}

\usepackage{graphicx}
\usepackage{epsfig}

\begin{document}

\title{FLUX LIMITER METHODS IN 3D NUMERICAL RELATIVITY}
\author{C. Bona, C. Palenzuela}
\maketitle
\begin{abstract}
New numerical methods have been applied in relativity to obtain a
numerical evolution of Einstein equations much more robust and
stable. Starting from 3+1 formalism and with the evolution
equations written as a FOFCH (first-order flux conservative
hyperbolic) system,  advanced numerical methods from CFD
(computational fluid dynamics) have been successfully applied. A
flux limiter mechanism has been implemented in order to deal with
steep gradients like the ones usually associated with black hole
spacetimes. As a test bed, the method has been applied to 3D
metrics describing propagation of nonlinear gauge waves. Results are
compared with the ones obtained with standard methods, showing a
great increase in both robustness and stability of the numerical
algorithm.
\end{abstract}

\section{Introduction}

From the very beginning, 3D numerical relativity has not been an
easy domain. Difficulties arise either from the computational side
(the large amount of variables to evolve, the large number of
operations to perform, the stability of the evolution code) or
from the physical side, like the complexity of the Einstein
equations themselves, boundary conditions, singularity avoidant
gauge choices, and so on. Sometimes there is a connection between
both sides. For instance, the use of singularity avoidant slicings
generates large gradients in the vicinity of black holes.
Numerical instabilities can be produced by these steep gradients.
The reason for this is that the standard evolution algorithms are
unable to deal with sharp profiles. The instability shows up in
the form of spurious oscillations which usually grow and break
down the code.

Numerical advanced methods from CFD (Computational Fluid Dynamics)
can be used to avoid this. Stable codes are obtained which evolve
in a more robust way, without too much dissipation, so that the
shape of the profiles of the evolved quantities is not lost. These
advanced methods are then specially suited for the problem of
shock propagation, but they apply only to strongly hyperbolic
systems, where one is able get a full set of eigenfields which
generates all the physical quantities to be evolved. In the 1D
case, these methods usually fulfill the TVD (Total Variation
Diminishing) condition when applied to transport equations. This
ensures that no new local extreme appear in the profiles of the
eigenfields, so that spurious oscillations are ruled out ab initio
(monotonicity preserving condition). Unluckily, there is no
general method with this property in the 3D case, mainly because
the eigenfield basis depends on the direction of propagation. We
will show how this can be achieved at least in some cases.

The specific methods we will use are known as flux limiter
algorithms. We will consider plane waves in 3D as a first
generalization of the 1D case, because the propagation direction
is constant. This specific direction leads then to an specific
eigenfield basis, so that the 1D numerical method can be easily
generalized to the 3D case.

The algorithm will be checked with a "Minkowski waves" metric. It can be
obtained by a coordinate transformation from Minkowski space-time. All the
metric components are transported while preserving their initial profiles. The
line element has the following form:
\begin{equation}\label{metric_minkowski}
    ds^2 = -H(x-t)\;dt^2 + H(x-t)\;dx^2 + dy^2 + dz^2
\end{equation}
where $H(x-t)$ is any positive function. We can choose a periodic
profile with sharp peaks so both the space and the time
derivatives of $H(x-t)$ will have discontinuous step-like
profiles. If we can solve well this case (the most extreme), we
can hope that the algorithm will work as well in more realistic
cases where discontinuities do not appear.

Minkowski waves are a nice test bed because the instabilities can
arise only from the gauge (there are pure gauge after all!). This
is a first step to deal with evolution instabilities in the
Einstein equations by the use of flux limiter methods. This will
allow us to keep all our gauge freedom available to deal with more
physical problems, like going to a co-rotating frame or adapting
to some special geometry. Advanced numerical methods take care of
numerical problems so that 'physical' gauge choices can be used to
take care of physics requirements.

\section{The system of equations}

We will use the well known 3+1 description of spacetime
\cite{Choquet48,Choquet62,ADM62} which starts by decomposing the
line element as follows:
\begin{equation}\label{metric}
    ds^2 = -\alpha^2\;dt^2 +
    \gamma_{ij}\;(dx^i+\beta^i\;dt)\;(dx^j+\beta^j\;dt) \;\;\;\;i,j=1,2,3
\end{equation}
where $\gamma_{ij}$ is the metric induced on the three-dimensional
slices and $\beta^i$ is the shift. For simplicity the case
$\beta^i=0$ (normal coordinates) will be considered. The intrinsic
curvature of the slices is then given by the three-dimensional
Ricci tensor $^{(3)}R_{ij}$, whereas their extrinsic curvature
$K_{ij}$ is given by:
\begin{equation}\label{Kij}
    \partial_t \gamma_{ij} = -2\:\alpha\:K_{ij}  \;
\end{equation}
In what follows, all the geometrical operations (index raising,
covariant derivations, etc) will be performed in the framework of
the intrinsic three-dimensional geometry of every constant time
slice. With the help of the quantities defined in
(\ref{metric},\ref{Kij}), the ten fourdimensional field equations
can be expressed as a set of six evolution equations:
\begin{eqnarray}\label{ADMKij}
   \partial_t K_{ij} = -\nabla_i\alpha_j
       + \alpha\; [{}^{(3)}R_{ij}-2K^2_{ij}
       + tr\:K\:K_{ij} - 8\pi\: (T_{ij}-\frac {T} {2} \gamma_{ij})]
\end{eqnarray}
plus four constraint equations
\begin{equation}\label{energy_constraint}
   ^{(3)}R - tr(K^2) + (tr\:K)^2 = 16 \pi \alpha^2 \: T^{00}
\end{equation}
\begin{equation}\label{momentum_constraint}
   \nabla_k\:{K^k}_{i} - \partial_i(tr\:K) = 8 \pi \alpha \: T^0_{\;i}
\end{equation}
The evolution system (\ref{ADMKij}) has been used by numerical
relativists since the very beginning of the field (see for
instance the seminal work of Eardley and Smarr
\cite{Eardley79}), both in spherically symmetric (1D) and
axially symmetric (2D) spacetimes. By the turn of the century, the
second order system (\ref{ADMKij}) has been rewritten as a
first-order flux conservative hyperbolic (FOFCH) system
\cite{PRL92,Choquet&York95,Frittelli&Reula96} in order to deal with the generic
(3D) case, where no symmetries are present. But the second order system
(\ref{ADMKij}) is still being used in 3D numerical calculations
\cite{Alcubierre01}, mainly when combined with the conformal
decomposition of $K_{ij}$ as introduced by Shibata and Nakamura
\cite{SN,BS}. In the system(\ref{Kij},\ref{ADMKij}) there is a
degree of freedom to be fixed because the evolution equation for
the lapse function $\alpha$ is not given. In the study of Black
Holes, the slicing is usually chosen in order to avoid the
singularity \cite{PRD97}:
\begin{equation}\label{lapse_evolution}
    \partial_t \ln \alpha = -\alpha\: Q  \;
\end{equation}
where:
\begin{equation}\label{Q}
   Q=f(\alpha)trK
\end{equation}
Three basic steps are needed to obtain a FOFCH system from the ADM
system. First, one must introduce some new auxiliary variables to
express the second order derivatives in space as first order.
These new quantities correspond to the space derivatives:
\begin{equation}\label{space_der}
    A_k = \partial_k \ln \alpha \;,\;\;\;
    D_{kij} = 1/2\: \partial_k \gamma_{ij} \;
\end{equation}
The evolution equations for these variables are:
\begin{eqnarray}\label{eqn_space_der}
    \partial_t A_k + \partial_k(\alpha\:f\:trK) &=& 0 \\
    \partial_t D_{kij} + \partial_k(\alpha\: K_{ij}) &=& 0
\end{eqnarray}
At the second step the system is expressed in a first order
balance law form
\begin{equation}\label{FOFCH}
    \partial_t \vec{u} + \partial_k F^k(\vec{u}) =
    S(\vec{u})  \;,
\end{equation}
where the array $\vec{u}$ displays the set of independent
variables to evolve and both "fluxes" $F^k$ and "sources" $S$ are
vector valued functions.
At the third step another additional independent variable is
introduced to obtain a strongly hyperbolic system \cite{PRD97}:
\begin{equation}\label{V}
    V_i = {D_{ir}}^{r} - {D^r}_{ri}
\end{equation}
and its evolution equation is obtained using the definition of
$K_{ij}$ from (\ref{Kij}) and switching space and time
derivatives in the momentum constraint
(\ref{momentum_constraint}). The result is an independent evolution
equation for $V_i$ while the previous definition (\ref{V}) in terms of space derivatives can be instead be considered as a first integral of the extended system. The extended array $\vec{u}$ will then contain the following 37
functions $\vec{u}=(\alpha$, $\gamma_{ij}$, $K_{ij}$, $A_i$,
$D_{kij}$, $V_i)$.

\section{The numerical algorithm}

Due to the structure of the equations, the evolution(represented
by the operator $E(\Delta t))$ described by (\ref{FOFCH}) can be
decomposed into two separate processes; the first one is a transport
process and the second one is the contribution of the sources.

The sources step (represented by the operator $S(\Delta t))$ does
not involve space derivatives of the fields, so that it consists
in a system of coupled non-linear ODE (Ordinary Differential
Equations):
\begin{equation}\label{Process_Sources}
    \partial_t \vec{u} =  S(\vec{u})  \;
\end{equation}
The transport step (represented by the operator $T(\Delta t)$)
contains the principal part and it is given by a set of
quasi-linear transport equations:
\begin{equation}\label{Process_Fluxes}
    \partial_t \vec{u} + \partial_k F^k(\vec{u})= 0  \;
\end{equation}
The numerical implementation of these separated processes is quite
easy. Second order accuracy in $\Delta t$ can be obtained by using the well known Strang splitting.
\begin{equation}\label{Strang}
      E(\Delta t) = S(\Delta t /2) \, T(\Delta t) \, S(\Delta t /2) \;
\end{equation}
According to (\ref{Kij},\ref{lapse_evolution}) the lapse and the
metric have no flux terms. It means that a reduced set of 30
quantities $\vec{u}=(K_{ij},A_i,D_{kij},V_i)$ are transported in
the second step over an inhomogeneous static background composed
by $(\alpha,\gamma_{ij})$. The equations for the transport step
(\ref{Process_Fluxes}) are given by:

\begin{eqnarray}\label{Fluxes}
    \partial_t K_{ij} + \partial_k(\alpha\: {\lambda^k}_{ij}) &=& 0 \\
    \partial_t A_k + \partial_k(\alpha\:f(\alpha)\:trK) &=& 0 \\
    \partial_t D_{kij} + \partial_k(\alpha\: K_{ij}) &=& 0 \\
    \partial_t V_k &=& 0
\end{eqnarray}
where:
\begin{equation}\label{Fluxes_K}
   {\lambda^k}_{ij}=D^k_{ij}-\frac{m}{2} V^k \gamma_{ij}
     + 1/2 \: \delta^k_i(A_j + 2 V_j - {D_{jr}}^r)
     + 1/2 \: \delta^k_j(A_i + 2 V_i - {D_{ir}}^r) \nonumber
\end{equation}
and m is an arbitrary parameter.

To evolve the transport step, we will consider flux-conservative
numeric algorithms \cite{LeVeque}, obtained by applying the balance to a
single computational cell. In the 1D case the cell goes from n to n+1 in
time ($t=n\cdot\Delta t$) and from j-1/2 to j+1/2 in space
($x_j=j\cdot\Delta x$), so that we have:
\begin{equation}\label{algorithm}
    U^{n+1}_j = U^n_j - \frac{\Delta t}{\Delta x}
              [F^{n+1/2}_{j+1/2}-F^{n+1/2}_{j-1/2}]
\end{equation}
Interface fluxes can be calculated in many different ways, leading to
different numerical methods. We will use here the well known MacCormack
method. This flux-conservative standard algorithm works well for smooth
profiles, as it can be appreciated in Figure 1.

\begin{figure}\label{sinus1d}
\centering
\hbox{\epsfig{file=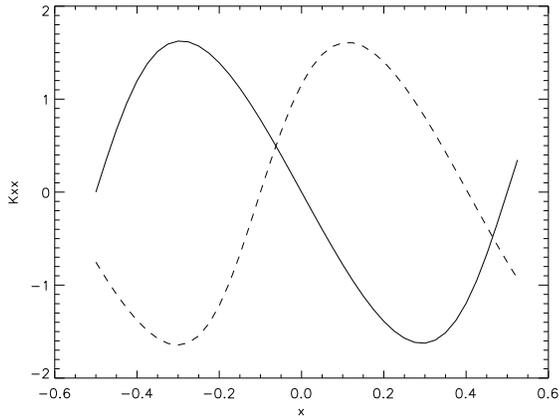,width=8cm}}
\caption{Plot
of $K_{xx}$ for the initial metric given by
(\ref{metric_minkowski}) with $H(x-t)=1+A\:\cos[\omega\:(x-t))]$
with periodic boundaries.Continuous line is the initial
condition.Dashed line is after 40 iterations}
\end{figure}

But this standard algorithm is not appropriate for step-like
profiles because it produces spurious oscillations near the steep
regions, as it can be appreciated in Figure 2.
\begin{figure}\label{step1d}
\centering
\hbox{\epsfig{file=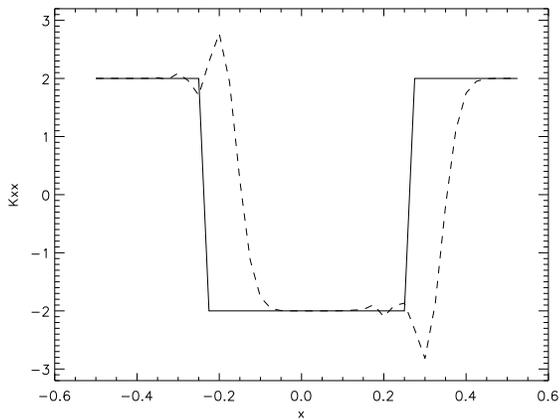,width=8cm}}
\caption{Same
as in Fig. 1 with the step-like initial data for $K_{xx}$.
Continuous line is the initial condition. Dashed line is after 10
iterations. Note the spurious oscillations around the corners}
\end{figure}

More advanced numerical methods must be used to eliminate (or at least
to reduce) these oscillations. These advanced methods use
information about the eigenfields and the propagation direction,
so the flux characteristic matrix along the propagation direction
must be diagonalized.

\section{Eigenfields}

We will use a convenient method to compute the eigenfields. Let us
study the propagation of a step-like discontinuity in the
transported variables $\vec{u}$ which will move along a specific
direction $n$ with a given velocity $v$. Information about the
corresponding eigenfields can be extracted from the well known
Rankine-Hugoniot shock conditions :
\begin{equation}\label{RHgeneral}
     v[u] = n_k[F^k(u)]
\end{equation}
where $[\;]$ represents the jump in the discontinuity. In our case
\begin{eqnarray}\label{RHconditions}
      v[K_{ij}] &=& n_r[\alpha \:{\lambda^r}_{ij}] \\
      v[A_k] &=& n_k[\alpha\: f(\alpha) \:trK] \\
      v[D_{kij}] &=& n_k[\alpha\: K_{ij}]  \\
      v[V_k] &=& 0
\end{eqnarray}
where we must note that both the background metric coefficients
and the propagation direction are supposed to be continuous, so
they are transparent to the $[\;]$ symbol.

If we develop this expressions we arrive at the following
conclusions, where $S_n = n^r\: S_r$ is the projection of the
quantity $S$ over $n$ and $S_{\bot}=S_k-S_n\: n_k$ are the
transverse components:

1) $[V_k],[A_{\bot}],[D_{\bot ij}]$ and $[A_n - f\: trD_n]$
propagate along $n$ with speed $v=0$. There are 18 such
eigenfields. For the line element given by (\ref{metric_minkowski}) $n^k$
is along the x axis and all these fields are actually zero.

2) $[{\lambda^n}_{ij}-tr\lambda^n n_i n_j]$ and $[K_{ij}-trK n_i
n_j]$ do generate eigenfields propagating along $n$ with speed
$v=\pm\alpha$ (light cones). There are only 10 such eigenfields
because all of them are traceless. For Minkowski waves, where
there is only gauge, all these combinations are zero. This
indicates that the correct way to get the traceless part of a
given tensor $S_{ij}$ in this context is just to take
$S_{ij}-trS\: n_i\: n_j$, so that the contribution of gauge modes
will disappear.

3) $[A_n]$ and $[trK]$ do generate eigenfields propagating along
$n$ with speed $v=\pm \sqrt{f} \alpha$ (gauge cones). There are 2
such eigenfields corresponding to the gauge sector. For Minkowski
waves, there are the only non-zero components. We are left with:
\begin{eqnarray}\label{RHgauge}
      v[trK] &=& \alpha [ A_n] \\
      v[A_k] &=& \alpha\: f(\alpha) [trK]\: n_k
\end{eqnarray}
so that $[A_k]$ is proportional to $n_k$. Now we can get the
gauge eigenfields:
\begin{equation}\label{gauge_eigenfield}
     \sqrt{f}\:n_k\: F^k(trK) \pm F(A_n)
\end{equation}

These eigenfields propagate along $n$ according simple advection
equations, a familiar situation in the 1D case. Although this
decomposition and diagonalization is trivial in 1D, it is very
useful in the multidimensional case for a generic direction $n$.

\section{Flux limiter methods}

The flux limiter methods \cite{LeVeque} we will use can be decomposed into
some basic steps. First of all the interface fluxes have to be calculated
with any standard second order accurate method (MacCormack in our
case). Then, the propagation direction $n$ and the corresponding
eigenfields can be properly identified at every cell interface.
Two advection equations (one for every sense of propagation) are
now available for the gauge eigenfluxes (\ref{gauge_eigenfield}).

Let us choose for instance the eigenflux which propagates to the
right (an equivalent process will be valid for the other eigenflux
propagating to the left). This interface eigenflux
$F^{n+1/2}_{j+1/2}$ can be understood as the grid point flux
$F^{n}_{j}$ plus some increment $\Delta_{j+1/2} =
F^{n+1/2}_{j+1/2} - F^{n}_{j}$. In general, the purpose of the
limiter is to use of a mixture of the increments $\Delta_{j+1/2}$
and $\Delta_{j-1/2}$ to ensure monotonicity. In our case we are
using a robust mixture which goes by applying the well known minmod rule to
$\Delta_{j+1/2}$ and $2 \Delta_{j-1/2}$. In that way, the limiter acts only in
steep regions, where the proportion between neighbouring increments exceeds a
factor of two. 

We can apply this method to the step-like initial data propagating
along the x axis. We can see in Figure 3 that the result is much
better than before. It can be (hardly) observed a small deviation
from the TVD condition, which is produced by the artificial
separation produced by the Strang splitting into transport and
non-linear source steps.

\begin{figure}\label{lstep1d}
\centering
\hbox{\epsfig{file=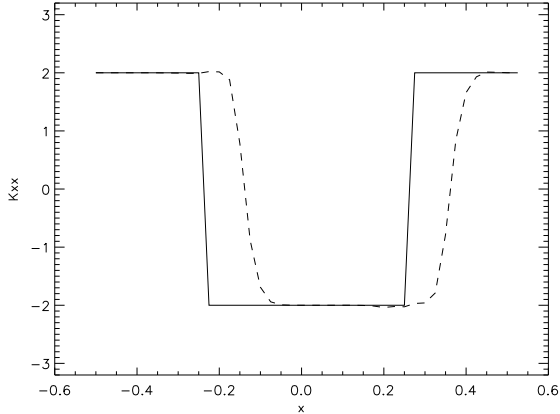,width=8cm}}
\caption{Same
as Fig.2 where the methods presented in this paper are applied.
Continuous line is the initial condition. Dashed line is after 10
iterations}
\end{figure}

This method can be applied, with the general decomposition described in section 4, to discontinuities which propagate along any
constant direction, and not only to the trivial case, aligned with
the x axis, that we have considered until now. To prove it, we have
rotated the metric of Minkowski waves in the x-z plane to obtain a
diagonal propagation of the profile. The line element in this case
has the following form:
\begin{eqnarray}\label{metric_2d}
   ds^2 = &-& H(\frac{x+z}{\sqrt{2}}-t)\;dt^2 +
          \frac{1}{2}[1+H(\frac{x+z}{\sqrt{2}}-t)]\;(dx^2+dz^2) +
          dy^2 \nonumber \\
          &+&  \frac{1}{2}[-1+H(\frac{x+z}{\sqrt{2}}-t)]\;(dx\:dz + dz\:dx)
\end{eqnarray}
We show the results in the Figure 4.
\begin{figure}\label{lstep2d}
\centering
\hbox{\epsfig{file=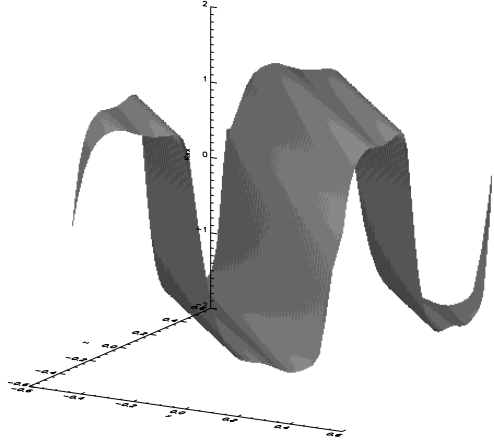,width=8cm}}
\caption{3D
plot of $K_{xx}$. The step-like profile has been propagated with
periodic boundary conditions until one full period (about 80
iterations in this case) has elapsed and it has returned to the
initial position}
\end{figure}
We can also see in Figure 5 a z=constant section of the same
results to allow a more detailed comparison with the initial data.

\begin{figure}\label{step2d_x}
\hbox{\epsfig{file=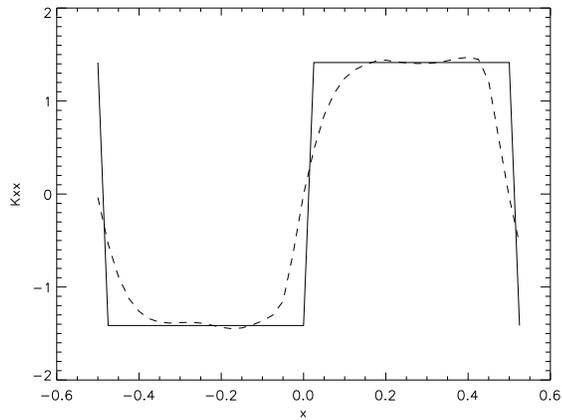,width=8cm}} \caption{Section
z=const. from Fig. 4. Continuous line is the initial condition.
Dashed line is after one full period}
\end{figure}

{\em Acknowledgements: This work has been supported by the EU Programme
'Improving the Human Research Potential and the Socio-Economic
Knowledge Base' (Research Training Network Contract (HPRN-CT-2000-00137)
and by a grant from the Conselleria d'Innovacio i Energia of the Govern 
de les Illes Balears}

\bibliographystyle{prsty}

\end{document}